\documentclass{kluwer}    

\newdisplay{guess}{Conjecture}

\usepackage[dvips]{graphicx}
\begin{document}
\begin{article}
\begin{opening}
\title{Constraints on extra-dimensions and variable constants from cosmological gamma
ray bursts}
\author{T. \surname{Harko}\thanks{email:harko@hkucc.hku.hk} and K. S. \surname{Cheng}\thanks{e-mail:hrspksc@hkucc.hku.hk}}
\runningauthor{Harko \& Cheng}\runningtitle{Constraints on
extra-dimensions from GRB's} \institute{Department of Physics, The
University of Hong Kong, Pok Fu Lam Road, Hong Kong}
\date{July 20, 2004}

\begin{abstract}
The observation of the time delay between the soft emission and
the high-energy radiation from cosmological gamma ray bursts can
be used as an important observational test of multi-dimensional
physical theories. The main source of the time delay is the
variation of the electromagnetic coupling, due to dimensional
reduction, which induces an energy dependence of the speed of
light. For photons with energies around 1 TeV, the time delay
could range from a few seconds in the case of Kaluza-Klein models
to a few days for models with large extra-dimensions. Based on
these results we suggest that the detection of the 18-GeV photon $
\sim $4500 s after the keV/MeV burst in GRB 940217 provides a
strong evidence for the existence of extra-dimensions. The time
delay of photons, if observed by the next generation of high
energy detectors, like, for example, the SWIFT and GLAST satellite
based detectors, or the VERITAS ground-based TeV gamma-ray
instrument, could differentiate between the different models with
extra-dimensions.
\end{abstract}

\end{opening}

\section{Introduction}

One of the most challenging issues of modern physics is the
possible existence of the extra-dimensions of the space-time
continuum. Multi-dimensional geometries are the natural framework
for the modern string/M theories \cite{Wi96} or brane models
\cite{Ho96}. String models also provide a natural and
self-consistent explanation for the possible variation of the
fundamental constants \cite{Ho79}. Hence the problem of the
extra-dimensions of the space-time continuum is closely related to
the problem of the variations of the fundamental constants, like,
for example the fine structure constant $\alpha $ or the speed of
light $c$. Most of the multi-dimensional theories contain a
built-in mechanism, which allows the variation of $\alpha $ and
$c$. The speed of propagation of particles in vacuum is modified
due to the supplementary effects induced by the extra-dimensions.

The confirmation that at least some gamma-ray bursts (GRBs) are
indeed at cosmological distances raises the possibility that
observations of these could provide interesting constraints on the
fundamental laws of physics. The study of short-duration photon
bursts, propagating over cosmological distances, is the most
promising way to probe the effects related to the existence of
extra-dimensions and quantum gravity effects. Data on GRBs may be
used to set limits on variations in the velocity of light. This
has been illustrated, by using BATSE and OSSE observations of the
GRBs that have recently been identified optically, and for which
precise redshifts are available, in \cite{Ma00}.

It is the purpose of the present paper to consider the effects of
the existence of the extra-dimensions on the propagation of high
energy photons. Based on the expressions for the photon time delay
we suggest that the detection of the 18-GeV photon $ \sim $4500 s
after the keV/MeV burst in GRB 940217 provides a strong evidence
for the existence of a multi-dimensional Universe.

The present paper is organized as follows. In Section II we review
the main physical mechanisms for TeV photon emission from GRB's.
The time delay of photons in multi-dimensional cosmological models
is considered in Section III. We discuss and conclude our results
in Section IV.

\section{TeV photons from gamma-ray bursts}

The mechanisms for TeV photon production in GRBs have been
reviewed recently by \cite{WaCh03}. One such mechanism is the
electron inverse Compton emission in GRB shocks, with an energy
spectrum that can be commonly described as $\nu F_{\nu }\propto
\nu ^{-p+1/2}$, with $2.05<p<2.9$. On the other hand, in the
reverse shocks the TeV spectrum is given by $\nu F_{\nu }\propto
\nu ^{-p/2+1}$, where $2<p<\sqrt{6}$.

Similar to the electrons, protons can also be accelerated up to
ultra-high energies higher than $10^{20}$ eV \cite{WaCh03},
producing a spectrum
characteristic of Fermi mechanism $dN_{p}/dE_{p}\propto E_{p}^{-p}$, where $%
E_{p}$ is the energy of the proton. These protons, accelerated in
both internal and external shocks can produce gamma rays with
energies of the order of TeV \cite{WaCh03}.
The protons can be accelerated up to $10^{20}-10^{21}$ eV for $%
\Gamma _{0}=10^{2}-10^{3}$ and therefore the energy of the
synchrotron photons can extend to the TeV band for $\epsilon
_{B}n\sim 1$, where $\epsilon _{B}$ is the shock energy carried
out by the magnetic field and $n$ is the number density of the
external medium. The energy spectrum from proton-synchrotron
radiation is $\nu F_{\nu }\propto \nu ^{(3-p)/2}$. If the spectrum
of the accelerated protons is that of the standard shock theory,
then $(3-p)/2\approx 0.5$ \cite{WaCh03}.

TeV gamma rays emitted from extra-galactic sources may collide
with diffuse cosmic infrared background photons, leading to
secondary $e^{+}e^{-}$ electron-positron pairs. The pair
production optical depth $\tau _{\gamma \gamma }$ depends on the
spectral energy distribution and the intensity of the cosmic
infrared background. Because of the high redshift of the
cosmological GRB sources, $\tau _{\gamma \gamma }$ also depends on
the evolution of the cosmic infrared background with the redshift.

For TeV $\gamma $-photons the pair production cross section is
maximized when the soft photon energy $\epsilon _{IR}$ is in the
infrared range. For a $1$ TeV gamma ray this corresponds to a soft
photon wavelength near the K-band ($\sim 2$ $\mu $m). Thus,
infrared photons with wavelengths around $2$ $\mu $m will
contribute most to the absorption of TeV gamma rays. Absorption of
$\gamma $-rays of energies below $\sim 15$ GeV is negligible
\cite{St92}.

In order to estimate the number of high-energy photons from a
gamma-ray burst we assume that the source gives a continuous
spectrum, from keV to TeV photons. The fluence (the energy per
unit area) $F_{\gamma }$ can be represented, as a function of the
photon energy $E_{\gamma }$, in the general form $F_{\gamma
}\left( E_{\gamma }\right) =CE_{\gamma }^{-\beta }\exp \left(
-\tau _{\gamma }\right)$, where we assume that $\beta $ is a
(model dependent) constant. The proportionality constant can be
obtained from the observational result that the observed fluence,
for $E_{\gamma }\sim 100$ keV, is of the order of $F_{\gamma
}\left( E_{\gamma }=100\textrm{ keV}\right) \sim 10^{-6}-10^{-4}$
erg cm$^{-2}$. Then the fluence can be represented as
\begin{equation}
F_{\gamma }\left( E_{\gamma }\right) =\left(
10^{-6}-10^{-4}\right) 10^{-7\beta }\left( \frac{E_{\gamma
}}{1\textrm{ TeV}}\right) ^{-\beta }e^{ -\tau _{\gamma }} \textrm{
erg cm}^{-2}\textrm{.}
\end{equation}

The photon number $N_{\gamma }$, which can be observed by a
detector with collection area $A_{col}$, is given by
\begin{equation}\label{num}
N_{\gamma }\left( E_{\gamma }\right) =1.602\times \left(
10^{-6}-10^{-4}\right) \times 10^{-7\beta }\times \left( \frac{E_{\gamma }}{1%
\textrm{ TeV}}\right) ^{-\beta -1}\times e^{ -\tau _{\gamma }}
\times A_{col}.
\end{equation}

By using Eq. (\ref{num}) one could estimate the number of TeV
photons that could be detected by using one of the next-generation
of ground based TeV gamma ray instruments, the VERITAS system. It
consists of  a seven, 12 m aperture telescopes, with six
telescopes located at the corners of a hexagon of side $80$ m
\cite{We03}. For TeV photons the photon collection area for this
system can be as large as $5\times 10^8$ cm$^2$. For the VERITAS
ground-based array the variation of $A_{col}$ is represented, as a
function of the photon energy $E_{\gamma }$, in Fig. 1.

\begin{figure}
\centerline{\includegraphics[width=12pc]{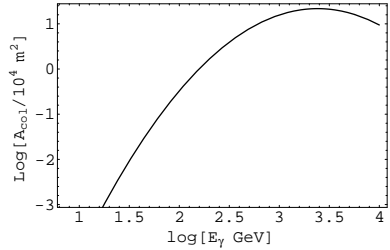}}
\caption{Variation, as a function of the photon energy $E_{\gamma
}$, of the photon collection area $A_{col}$ for the VERITAS
ground-based TeV gamma-ray instrument.}
\end{figure}

The number of high energy TeV photons is represented, for three
different gamma ray emission models and for two optical depth
models (corresponding to the lower and upper limit of the infrared
extra-galactic background, respectively), in Figs. 2 and 3.

\begin{figure}
\centerline{\includegraphics[width=12pc]{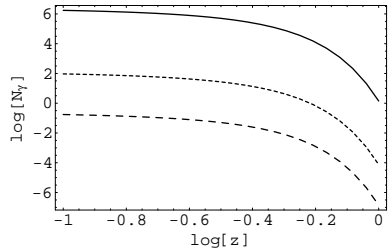}}
\caption{Variation of the photon number $N_{\gamma }$, for a
$E_{\gamma }=1$ TeV photon energy, as a function of the redshift
$z$ (in a logarithmic scale) for different high energy TeV
gamma-ray production models: proton synchrotron radiation ($\beta
=-0.5$) (solid curve), electron inverse Compton scattering
emission in internal shocks ($\beta =0.11$) (dotted curve) and
gamma ray emission via inverse electron Compton scattering in
external forward shocks ($\beta =0.5$) (dashed curve). The optical
depth for pair production is calculated by considering the lower
limit for the extra-galactic infrared field density. The assumed
collection area of the detector is of the order $A_{col}=5\times
10^8$ cm$^2$.}
\end{figure}

\begin{figure}
\centerline{\includegraphics[width=12pc]{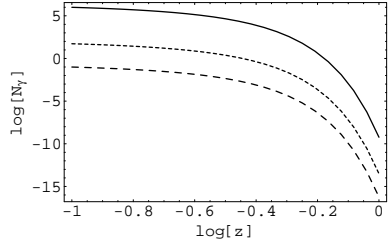}}
\caption{Variation of the photon number $N_{\gamma }$, for a
$E_{\gamma }=1$ TeV photon energy, as a function of the redshift
$z$ (in a logarithmic scale) for different high energy TeV
gamma-ray production models: proton synchrotron radiation ($\beta
=-0.5$) (solid curve), electron inverse Compton scattering
emission in internal shocks ($\beta =0.11$) (dotted curve) and
gamma ray emission via inverse electron Compton scattering in
external forward shocks ($\beta =0.5$) (dashed curve). The optical
depth for pair production is calculated by considering the upper
limit for the extra-galactic infrared field density. The
collection area of the detector is of the order $A_{col}=5\times
10^8$ cm$^2$.}
\end{figure}

Hence, as one can see from Figures 2 and 3, the large collection
area, which for the VERITAS-4 array is of the order of $2.2\times
10^5$ m$^2$ for $10$ TeV photons \cite{We03a}, makes possible the
detection of high energy photons from GRBs with redshift $z$
smaller than $0.3-0.5$.

\section{Photon delay in multi-dimensional universes}

The energy-dependence of the speed of light of the photon due to
the presence of an extra-dimension is given by \cite{Ha04}
\begin{equation}
c=c_{0}\left[1+\varepsilon\left( \frac{E}{E_{K}}\right) ^{\beta
}\right], \label{eq7}
\end{equation}
where we denoted $E_{K}=c^{4}\Delta v/G$, with $\Delta v=v-v^{0}$
describing the variation of the size of the fifth dimension
between the moments of the emission and detection of a photon.
$\varepsilon=\pm 1$ is a constant, related to the sign of the
fifth dimension.

In the case of isotropic homogeneous cosmological models with
large non-compact extra-dimensions there is a non-zero
contribution in the four-dimensional space-time (the brane) from
the $5$-dimensional Weyl tensor from the bulk, expressed by a
scalar term $U$, called dark radiation \cite{Ch02}. The ``dark
radiation'' term is a pure five dimensional effect.

By taking into account the expressions for the variation of the
speed of light we obtain the following general equation describing
the time delay of two photons with different energies \cite{Ha04}:
\begin{equation}\label{1}
\Delta t=H_{0}^{-1}f^{(\beta )}\left( E_{1},E_{2}\right) \int_{0}^{z}\frac{%
\left( 1+z\right) ^{\beta -1}}{\sqrt{\Omega _{\Lambda }+\Omega
_{M}\left( 1+z\right) ^{3}+\Omega _{U}\left( 1+z\right) ^{4}} }dz,
\end{equation}
where $H_{0}=72$ km s$^{-1}$ Mpc$^{-1}$, $\Omega _{M}\approx 0.3$ and $%
\Omega _{\Lambda }\approx 0.7$ are the mass density parameter and
the dark energy parameter, respectively. $\Omega _{U}$ is the dark
radiation parameter and the function $f^{(\beta )}\left(
E_{1},E_{2}\right)$ describes the different physical models
incorporating extra-dimensional and/or quantum gravitational
effects. For $\beta =1$ we have the linear model, with
$f^{(1)}\left( E_{1},E_{2}\right)=\Delta E/E_K^{(1)}$, where we
denoted $\Delta E=E_{1}-E_{2}$.

The linear model corresponds, from the point of view of the
extra-dimensional interpretation, to an Einstein-Yang-Mills type
model. A linear energy dependence of the difference of the photon
velocities has also been considered in \cite{El03}.

A quadratic model of the form $f^{(2)}\left(
E_{1},E_{2}\right)=\left(\Delta E/E_K^{(2)}\right)^2$, can also be
considered in quantum gravitational models, in which selection
rules, such as rotational invariance, forbid first order terms
\cite{El03}. The function $f^{(3)}\left(
E_{1},E_{2}\right)=\left(E_{1}^{3}-E_{2}^{3}\right)/E_K^{(3)}$,
corresponding to $\beta =3$, describes the effect of a pure
five-dimensional gravity on the propagation of light in
four-dimensions. In the above equations we denoted by $E_K^{(\beta
)}$, $\beta =1,2,3$ the energy scales associated with the
different types of extra-dimensional and/or quantum gravity
mechanisms.

If the effects of extra-dimension do exist, then the detected time
profiles between the KeV/MeV/GeV burst and the TeV burst should be
very different. The comparison of the time profiles at the emitter
and observer is presented in Fig. 4.

\begin{figure}
\centerline{\includegraphics[width=14pc]{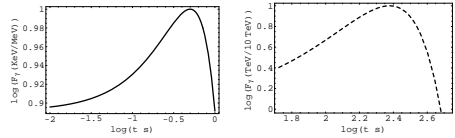}}
\caption{Comparison, in arbitrary units, of the initial KeV/MeV
time profile of the GRB emission occurring at a redshift $z=3$
(assumed to have a Gaussian form), with a duration of $\tau =1$ s
(solid curve), and the TeV time profile at the detector, modified
due to the presence of multi-dimensional effects, for the linear
model, with the fundamental energy scale $E_K=1.2\times 10^{19}$
GeV (dashed curve). For the mass and dark energy parameters we
have used the values $\Omega _{M}=0.3$ and $\Omega _{\Lambda
}=0.7$, respectively.}
\end{figure}

\section{Discussions and final remarks}

By using Eq. (\ref{1}) we can evaluate the extra-dimensional and
/or quantum gravity energy scale, which follows from the time
delay of the $18$ GeV photon in GRB 940217 \cite{Hu94}. GRB 940217
is a very strong burst, with a total fluence above $20$ keV of
$(6.6\pm 2.7)\times 10^{-4}$ erg cm$^{-2}$ and a duration of $\sim
$180 s in the BATSE range. After the low-energy emission has
ended, an $18$ GeV energy photon and 36 photons with 137 MeV
energy have been recorded, after $\sim $5400 s following the low
energy emission. Assuming that the photon originated at $z\sim 2$
and taking $\Delta t\approx 4500\pm800$ s, where we have also
included the uncertainty in the moment of photon emission during
the burst, we obtain for the linear energy scale the value
$E_K^{(1)}\approx \left(1.75-2.51\right)\times 10^{15} {\rm GeV}$.

This value is very close to the value $E_K^{(1)}\approx 6.9\times
10^{15}$ GeV obtained by using a wavelet technique analysis of
BATSE and OSSE data \cite{El03}. For the quadratic model energy
scale we have $E_K^{(2)}\approx \left(1.77-2.12\right)\times
10^{8} {\rm GeV}$.

In the case of quadratic quantum gravity corrections the
corresponding energy scale, derived by using BATSE and OSSE data,
is $E_K^{(2)}\geq 2.9\times 10^6$ GeV \cite{El03}. For the pure
Einstein gravity five-dimensional model we obtain
$E_K^{(3)}\approx \left(8.28-9.34\right)\times 10^{5} {\rm GeV}$.

From the point of view of multi-dimensional theories, the linear
model could describe the effects on the propagation of light in
Randall-Sundrum or Einstein-Yang-Mills type models \cite{Ha04}.
The quadratic model is specific for quantum gravity effects, while
the $\beta =3$ case could describe the case of pure Einstein
gravity in five dimensions. In all these cases the delay of the
$18$ GeV photon fixes the corresponding energy scales.

The remarkable concordance between the linear quantum
gravity/extra-dimensional energy scale obtained from the present
study of the time delay of the $18$ GeV photon in GRB 940217 and
from the independent study of the BATSE and OSSE data \cite{El03}
strongly suggest that this time delay could be the signature of
the extra-dimensions or/and quantum gravitational effects.

There are several proposed, satellite based GRB research projects.
SWIFT, a multi-wavelength GRB observatory will be launched in
early 2004. It has the optimum capabilities in determining the
origin of GRBs and their afterglows, providing redshifts for the
bursts and multi-wavelength light curves. A wide-field gamma-ray
camera will detect approximately 1000 GRBs in 3 years with a
sensitivity 5 times fainter than the BATSE detector aboard Compton
GRO. Sensitive narrow-field X-ray and UV/optical telescopes will
be pointed at the burst location in 20 to 70 seconds by an
autonomously controlled "swift" spacecraft. SWIFT will acquire
high-precision locations for gamma ray bursts and will rapidly
relay a 1-4 arc-minute position estimate to the ground within 15
seconds (http://swift. gsfc. nasa. gov, 2003).

Another satellite based experiment, the Gamma-ray Large Area Space
Telescope (GLAST), a high energy ($30$ MeV-$300$ GeV) gamma-ray
astronomy mission, is planned for launch at the end of 2006
\cite{We03, We03a}. GLAST should detect more than 200 GRBs per
year, with sensitivity to a few tens of GeV for a few bursts.
GLAST could also detect the energy- and distance-dependent
dispersion of ($10$ ms / GeV / Gpc), predicted by quantum gravity
and multi-dimensional models within 1 - 2 years of observations
\cite{Ha04}.

The possibility that the very high energy component of the
gamma-ray bursts might be delayed makes the detection with the
sensitive ground based atmospheric telescopes more feasible. The
time delay between the registration of the MeV burst and its
notification to the ground-based observer can be as little as ten
seconds. The slew times for the large telescopes varies from 20
seconds to 5 minutes, but can be much less if the observing target
for the telescope happens to be close to the gamma ray source
direction. In principle the telescope can continue to monitor the
source position for up to eight hours (depending on its elevation
and the position of Sun and Moon) and on succeeding nights.

The sensitivity that can be achieved is, for the current Whipple
10 m telescope of the order of $8\times 10^{-8}$ ergs/cm$^2$ for a
duration of the burst of $1$ and $10$ seconds, and of $2.4\times
10^{-7}$ ergs/cm$^2$ for a 100 s duration burst. For the proposed
VERITAS array of telescopes \cite{We03} the minimum fluence is
$10^{-8}$ ergs/cm$^2$ for $1$ and $10$ second bursts and $3\times
10^{-8}$ ergs/cm$^2$ for $100$ s bursts.

These data can be regarded as representative of the present
generation of operating telescopes and for those that will come
on-line the next few years (CANGAROO III, HESS, MAGIC and
VERITAS). Delayed gamma-ray bursts will be also dispersed, since
the velocity of light is a function of energy; hence the energy
resolution of the detectors will be critical. The minimum
detectable fluence is limited by the duration of the burst. For
short times the fluence is limited by the number of photons, since
there is virtually no background \cite{We03a}.

Therefore, the detection of the time delay between Tev and GeV/MeV
photons from GRBs could represent a new possibility for the study
and understanding of some fundamental aspects of the physical laws
governing our universe.


\end{article}
\end{document}